\documentclass[conference]{IEEEtran}

\ifCLASSINFOpdf
\usepackage[pdftex]{graphicx}
\else
\fi

\usepackage{cite}
\usepackage{graphicx}
\usepackage{mathtools}
\usepackage{float}
\usepackage{amsmath}
\usepackage[utf8]{inputenc}
\usepackage[final]{pdfpages}
\usepackage{pdfpages}
\usepackage{lipsum}
\usepackage{textcase}

	\pdfoutput=1

\begin{document}

\title{UWB Channel Sounding and Modeling for UAV Air-to-Ground Propagation Channels }

\author{\IEEEauthorblockN{Wahab Khawaja\IEEEauthorrefmark{2},
Ismail Guvenc\IEEEauthorrefmark{1}\IEEEauthorrefmark{2},
David Matolak\IEEEauthorrefmark{3}~\IEEEmembership{Fellow,~IEEE}}
\IEEEauthorblockA{\IEEEauthorrefmark{1}Department of Electrical and Computer Engineering, Florida International University, Miami, FL}
\IEEEauthorblockA{\IEEEauthorrefmark{2}Department of Electrical and Computer Engineering, North Carolina State University, Raleigh, NC}
\IEEEauthorblockA{\IEEEauthorrefmark{3}Department of Electrical and Computer Engineering, University of South Carolina, Coloumbia, SC}
Email: \{wkhaw001, iguvenc\}@ncsu.edu, matolak@cec.sc.edu
	
}

\maketitle
\begin{abstract}
Unmanned aerial vehicles (UAVs) are expected to be used extensively in the near future in applications such as aerial surveillance, transportation, and disaster assistance. The conditions under which UAVs operate are different from those of conventional piloted aircrafts. This necessitates development of new air-to-ground (AG) propagation channel models for UAVs. To our best knowledge, there are limited studies in the literature on sounding and modeling of ultrawideband (UWB) AG propagation channels. In this work, comprehensive UWB measurements are conducted for various UAV communication scenarios using Time Domain P410 UWB kits. Both time and frequency domain analysis of the measured data are carried out. Based on the measured data, stochastic path loss and multipath channel models are developed to characterize AG UWB propagation channels. 


\end{abstract}

\IEEEpeerreviewmaketitle

\section{Introduction}

There has been an exceptional interest in commercial unmanned aerial vehicles (UAVs) during the past several years, which find applications in areas such as filming, entertainment, disaster relief, agriculture, construction management, and communications. According to market research firm Tractica, commercial UAV shipments are expected to rise drastically from 80,000 units in 2015 to 2.7~million units in 2025, and services enabled by commercial UAVs will rise from \$170 million in 2015 to \$8.7 billion within the next 10~years~\cite{Tractica}. Google and Facebook have been recently investigating the use of unmanned aerial platforms to deliver Internet connectivity in rural areas. UAVs can also be used to deliver broadband wireless connectivity to hot spot areas during temporary events, and during emergencies/disasters such as earthquakes when the existing communications infrastructure can get  damaged~\cite{Arvind_VTM_2016}.

This emerging interest in UAVs necessitates studying propagation characteristics for air-to-ground (AG) UAV channels for link budget analysis and system design. To this end, ultrawideband (UWB) signals~\cite{FCC1,guvenc2008ultra,guvenc2011} allow to capture multipath components (MPCs) with a fine temporal resolution, which makes UWB an appealing technology for developing wideband propagation models. Large bandwidth of UWB can also facilitate high data rates, better penetration through materials, and co-existence with narrow band networks for UAV AG communications. To our best knowledge, there are no comprehensive and dedicated UWB channel models for UAV AG propagation channels. Current UWB propagation channel models developed for other scenarios~\cite{Molisch1,Molisch01,new1} can not be applied to the UAV AG channels due to different propagation environments.




\begin{figure}[!t]
	\centering
	\includegraphics[width=0.8\columnwidth]{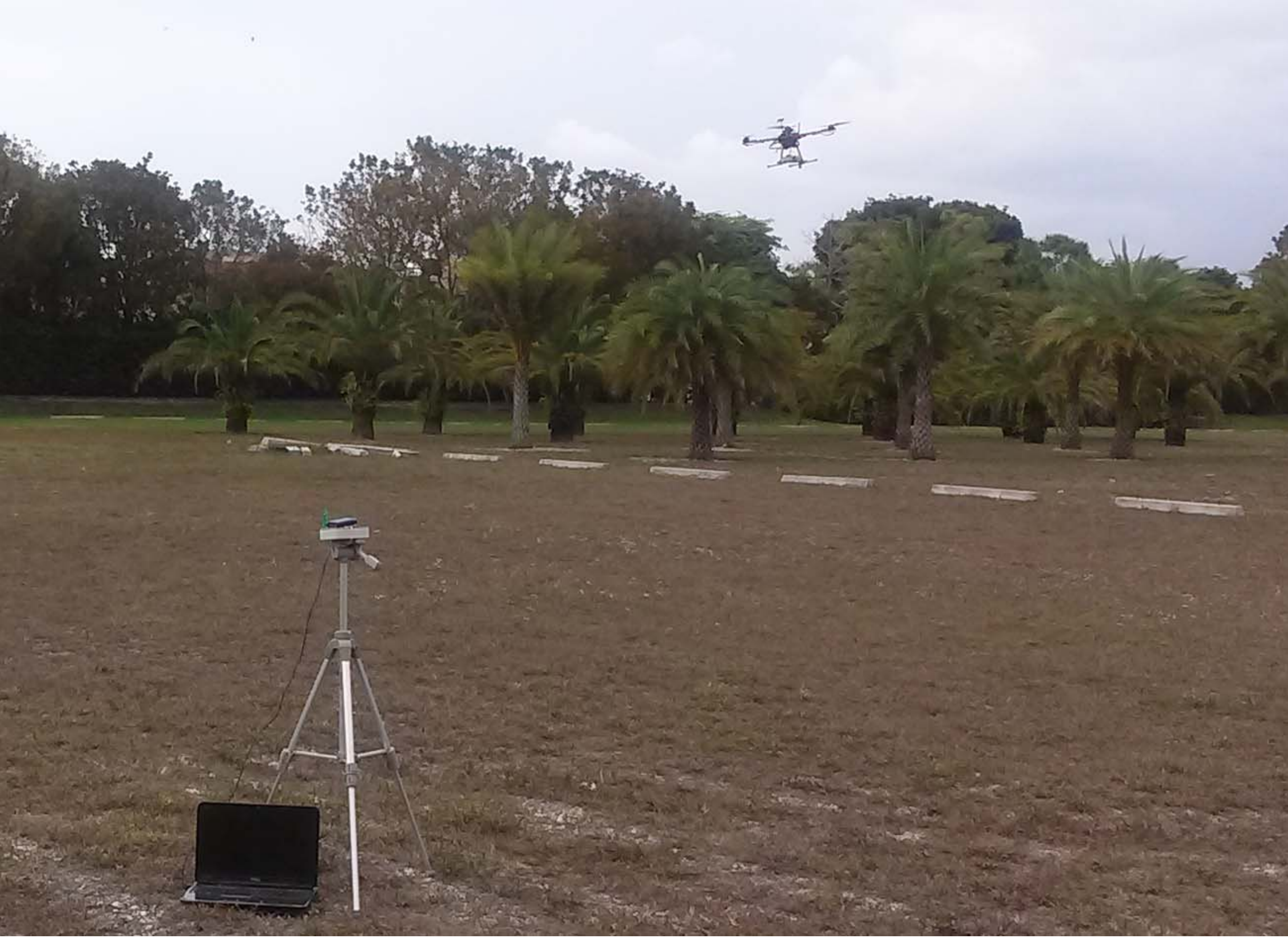}
	\vspace{-0.2cm}
	\caption{Experimental setup for UAV AG propagation measurements.}\label{Fig:Exp_Layout_actual}
	\vspace{-0.5cm}
\end{figure}

While UAV propagation channels have been studied in the literature, most existing work focus on addressing path loss characteristics of AG channels. For example, a ray tracing based path loss model for AG channels has been proposed in \cite{Qix} for urban areas, which is dependent on the elevation angle between the flying platform and the ground station. In~\cite{Hourani}, a statistical propagation channel model for AG path loss has been proposed for urban environment. Again, the prediction is dependent on the elevation angle between airborne transmitter and receiver on ground as well as properties of the urban environment, while it lacks corroborating measurement results. A geometric channel model for AG communication has been proposed in~\cite{Newhall}, which is applicable for both narrow and wideband channels. The model can be used to obtain amplitude, delay, time difference of arrival, and phase of the arriving MPCs, and provides spatio-temporal characteristics of signals for multiple array antennas. In~\cite{Motlak_new}, AG channel model characterization has been performed using UAV based measurements over water in L and C frequency bands. The measurement results are used to generate path loss and wideband dispersive channel models based on tapped delay line model. In \cite{new2} marine channel model is carried out using UAVs employing the ray tracing.

Contribution of this work can be summarized as follows: 1) to report our UAV AG channel measurements (see Fig.~\ref{Fig:Exp_Layout_actual}) for the $3.1$--$5.3$~GHz UWB spectrum, and 2) to develop statistical models to characterize the large scale fading, multipath propagation, and small scale fading in various scenarios based on the measured data. A comprehensive measurement campaign was carried out in open ground area and a sub-urban area in a variety of channel conditions. Our results show that the proposed stochastic model closely models the empirical UAV AG data.

\section{UWB Channel Sounding for UAVs}\label{Sec:UWB_Chn_Sound}

In this section, we will first describe our channel sounding procedure with Time Domain P410 UWB kits. Subsequently, we will describe the UAV channel sounding experimentation for various AG scenarios.


\subsection{Channel Sounding with P410 Kits}

For UWB channel sounding, Time Domain P410 UWB radios are used in bi-static mode~\cite{wahab}. In this mode a transmitter radio sounds the channel by sending short duration pulses at regular intervals of time. The transmitted pulse repetition rate is $10.1$~MHz. There is no need for physical synchronization between the transmitter and the receiver. A rake receiver is used for collection of MPCs at the receiver with an adjustable sampling rate. The frequency of operation for the P410 UWB kits is from $3.1$~GHz to $5.3$~GHz with an operational center frequency of $4.3$~GHz. The maximum transmit power is limited to $-14.5$~dBm from P410 radios, which falls within the FCC spectral mask. The antennas used in the experiment are BroadSpec UWB planar elliptical dipole antennas. The amplitude response of the antennas over the band is approximately flat. 

Clean algorithm \cite{clean} is used for obtaining refined channel impulse response (CIR). Fig.~\ref{Fig:pulse}(a) shows the normalized amplitude of the raw received pulses in blue. The waveforms in red are the reconstructed ones by convolving the CIR in the Fig.~\ref{Fig:pulse}(b) with the template waveform. The difference is due to the imperfections in the clean algorithm. Fig.~\ref{Fig:pulse}(b) is the CIR of the received waveform obtained by performing deconvolution of the received waveform with template waveform. The dashed blue lines show the threshold that is being set at $10\%$ of input signal, where all CIR samples below the threshold are discarded.  

\begin{figure}[!t]
	\centering
	\vspace{.7cm}
\includegraphics[width=\columnwidth]{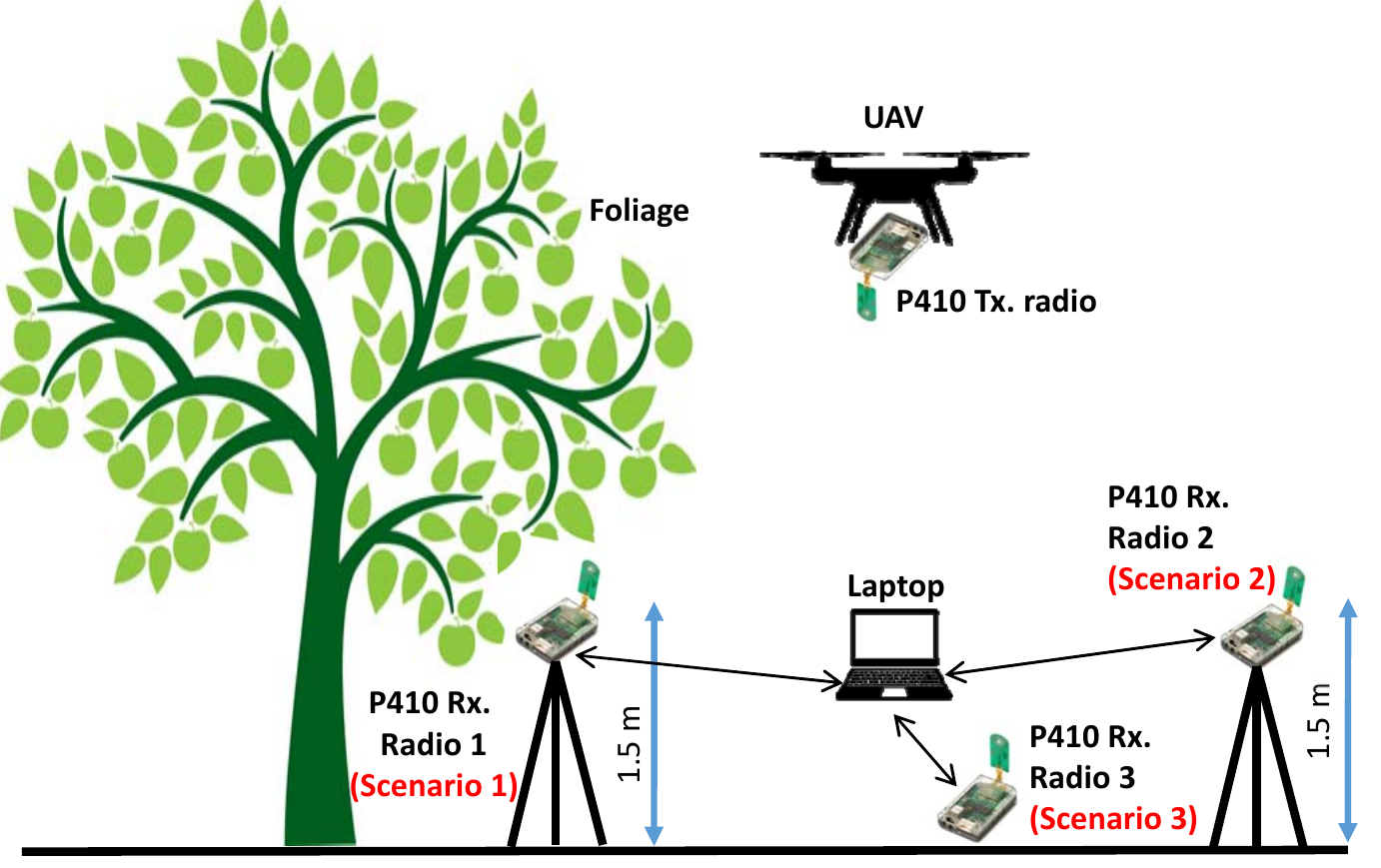}
	\caption{Layout for the UAV channel sounding scenario.}\label{Fig:Exp_Layout}
\end{figure}

\begin{figure}[!h]
	\centering
	\vspace{.5cm}
\includegraphics[width=\columnwidth]{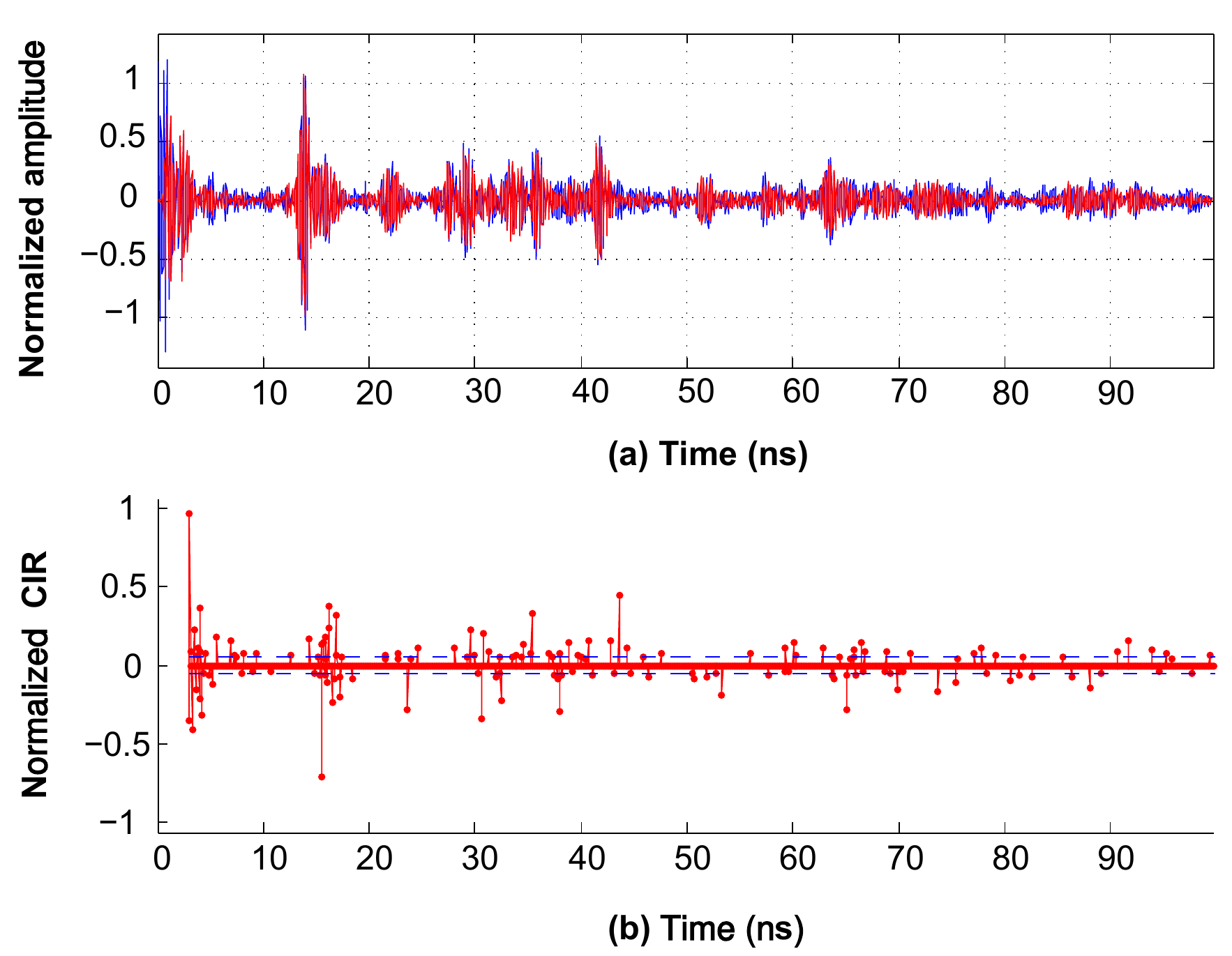}
	\caption{(a) Normalized amplitude of pulses with respect to time (blue represents received pulses and red represents reconstructed pulses), (b) CIR with respect to time.}\label{Fig:pulse}
\end{figure}

\subsection{Layout for UAV Channel Measurements}

The UAV channel propagation measurements were carried out at Florida International University Campus using the P410 UWB kits. The transmitter radio is placed on Tarot~650 quad-copter UAV belly such that the antenna is vertically facing down. In this setting, the beam pattern in the azimuth plane is in the form of circles spreading outward that can provide optimum coverage. The receiver is placed at two different heights from the ground.

In UAV based AG channel sounding, the use of very high platforms from ground for UAVs are rare~\cite{vacek}. Therefore, in our experiments, we varied the height of the UAV from $4$~m to $16$~m in steps of $4$~m. We considered  three scenarios each in open and sub-urban areas, all with line-of-sight (LOS) communication.
\begin{itemize}
\item  Scenario~1: The terrestrial receiver is at height of $1.5$~m from ground and in foliage. We placed the radios under the tree, such that branches and leaves were acting as foliage that hamper the direct LOS.
\item Scenario~2: The receiver is at a height of $1.5$~m from the ground, no foliage.
\item Scenario~3: The receiver is placed at $7$~cm from ground. 
\end{itemize}
A layout of three experimentation scenarios is shown in Fig.~\ref{Fig:Exp_Layout}.

In the experiments, the distance between the transmitter and receiver is adjusted and calculated using the Global Positioning System (GPS) coordinates of each node. Due to the small link distance the ellipsoidal effect of earth is negligible. Therefore all the calculations for distance from GPS coordinates were based on the spherical earth. The raw data from measurements is in the form of CIRs, which are later processed in Matlab. For the path loss analysis, we obtained the data with the UAV in motion and the UAV hovering at a fixed position for the three scenarios. The height of the UAV is also changed to four different distances.


\section{UWB Channel Modeling for UAVs}

Based on the collected measurement data as described in Section~\ref{Sec:UWB_Chn_Sound}, the characterization of AG channel model is divided into two main parts. In the first part, large scale fading is covered, which includes path loss and shadowing. In the second part, power delay profiles (PDPs) and small scale fading are considered based on the measured CIRs. The auxiliary parameters of the channel such as the mean excess delay and the root mean square delay spread (RMS-DS) are calculated from the measured PDPs.

\subsection{Large Scale Parameters}

For large scale parameter modeling, only LOS measurements for the AG channel were taken. The value of the distance between the transmitter and the receiver was calculated from each node's GPS longitude and latitude reading. The reference distance is taken as $d_{\rm 0} = 1$~m. The relative orientation of UAV in the azimuth direction to the ground nodes is considered to be irrelevant as the antennas are omni-directional and taken as zero. A modified free space path loss model (dB) in terms of both link distance and height is proposed based on the measured values when the UAV is assumed to be static:
\begin{align}
PL(d)&=PL_{\rm 0}+10 \alpha\log_{10}(d/d_{\rm 0})  \nonumber\\
        &- 10\log_{10}(\Delta h/h_{\rm opt}) +10\log_{10}c_{\rm p} + S,\label{Eq:1}
\end{align}
where $PL_{\rm 0}$ (in dB) is the reference path loss corresponding to reference distance $d_{\rm 0}$, $\alpha$ is the path loss exponent, $\Delta h = |h_{\rm gnd}-h_{\rm opt}|$, $h_{\rm gnd}$ is the height of the receiver above the ground, $h_{\rm opt}$ (can vary for different environments) is the minimum height of the receiver that gives the lowest path loss, $C_{\rm p} = 10\log_{10}c_{\rm p}>=0\ dB$ is the constant loss factor due to foliage and losses resulting from antenna orientations on UAV, and $S\sim\mathcal{N}(0,\sigma^2)$ is the shadowing variable, which is a zero mean Gaussian random variable with standard deviation $\sigma$.

The variations in the path loss will be due to the link distance between the transmitter and receiver and the clutter around it. Other minor factors that may influence path loss include the type of the antenna, its orientation on the UAV, and the type of the UAV, which are considered to be negligible in our case.

As the motion of UAV introduces Doppler, the effect of frequency dependence on the path loss cannot be ignored \cite{molisch00}. The effect of the frequency change will be due to the Doppler effect, and~\eqref{Eq:1} can be modified as follows:
\begin{align}
\begin{split}
PL(d) &=PL_{\rm 0} + 10\alpha\log_{10}(d/d_0) - 10\log_{10}\big((\Delta h)/h_{\rm opt}\big) \\
   &+10\log_{10}c_{\rm p} + 10x\log_{10}((f_{\rm e}+\Delta f)/f_{\rm e}) + S,\label{Eq:2} \end{split}
         \end{align}
where $\Delta f = (\Delta v/c)f_{\rm e}$, is the Doppler variation in the frequency due to the speed $v$ of the UAV relative to the receiver on the ground, $f_{\rm e}$ is the emitted frequency, $f= f_{\rm e}+\Delta f$  is the observed frequency at the receiver, and $x$ is the frequency dependence factor of path loss. At small velocities of few 10's of m/s, the factor $10x\log_{10}((f_{\rm e}+\Delta f)/f_{\rm e})$ is essentially negligible.

The measured average path loss (in dB) as a function of distance for UAV AG channels is obtained as follows~\cite{Merwaday}
\begin{align}
PL(d) &= PL(d_{\rm 0}) + 10\log_{10} \frac{\sum_{\forall i}P_{\rm d_0}[i]}{\sum_{\forall i}P_{\rm d}[i]},\\
P_{\rm d}[i] &= \sum_{k=1}^{N}|h[i,k]|^2/N_{\rm tot}, \  \ ~i = 1,2,...T~,
\end{align}
where $N = 1,2,...N_{\rm tot}$, and $P_{\rm d}[i]$ is the average PDP at distance $d$ with respect to time instants $i$ obtained by averaging over all the scans $N$, 
and $\sum_{\forall i}P_{\rm d_0}[i]$, $ \sum_{\forall i}P_{\rm d}[i]$ gives the total energies from all MPCs at distances $d_{\rm 0}$ and $d$. Each scan $N$ has a fixed time bin of $T$ seconds and $i$ represents the time index of the samples in the bin. $N_{\rm tot}$ is the total number of scans. Each scan time bin captures samples at a sampling rate of $T_{\rm s}$. The total number of distinct samples in each scan bin are $T/T_{\rm s}$. In our case we have taken 25 scans or CIRs for each scenario, and therefore we have $N_{\rm tot} = 25$. The value of $T_{\rm s}$ is $0.06$~ns, while the value of $T$ is $100$~ns.

\subsection{Channel Impulse Response}

In this paper, we model the CIR $h(t)$ using the Saleh Valenzuela channel model~\cite{Saleh}
\begin{equation}
h(t) = \sum_{n=0}^{N} \sum_{m=0}^{M} a_{n,m}(t)\exp(j\phi_{n,m}(t))\delta(t-\Gamma_{n}-\tau_{n,m}(t)),\label{Eq:CIR}
\end{equation}
where $N$ is the total number of clusters, $M$ is the total number of MPCs per cluster, $a_{n,m}(t)$, $\phi_{n,m}(t)$, $\tau_{n,m}(t)$ represents the amplitude, phase and delay of the $m^{\rm th}$ MPC in the $n^{\rm th}$ cluster as a function of time, $\Gamma_n$ is the delay of the $n^{\rm th}$ cluster. The phase is considered to be uniformly distributed random variable between the interval $[0,2\pi]$, and therefore it is neglected. In UWB channel the amplitude and delay terms vary slowly with respect to time and can be considered as time invariant. Therefore, we can rewrite~\eqref{Eq:CIR} as follows
\begin{equation}
h(t) = \sum_{n=0}^{N} \sum_{m=0}^{M}a_{n,m}\delta(t-\Gamma_{n}-\tau_{n,m})~.
\end{equation}
The distribution of arrival times for clusters and MPCs within each cluster is modeled as Poison process with respective cluster and ray arrival rates represented as $\Lambda$ and $\lambda$, respectively. Then, the distribution of cluster and ray arrival times can be written as~\cite{Saleh}
\begin{align}
p(\Gamma_n|\Gamma_{n-1}) &= \Lambda \exp(-\Lambda(\Gamma_n-\Gamma_{n-1}))~,\\
p(\tau_{n,m}|\tau_{n,m-1}) &= \lambda \exp(-\lambda(\tau_{n,m}-\tau_{n,m-1}))~.
\end{align}

The PDP within each non-overlapping cluster can be obtained as follows~\cite{molisch00}
\begin{equation}
P_{n}(t) = E(a_{n,m}^2)\exp(-\tau_{n,m}/\beta_n)\delta(t-\tau_{n,m}),
\end{equation}
where $P_{n}(t)$ is the PDP for $n$th non-overlapping cluster, $E$ stands for expected value, $E(a_{n,m}^2)$ represents the average power due to $m$ MPCs of the $n^{\rm th}$ cluster, and $\beta_{n}$ is the intra cluster decay constant of $n^{\rm th}$ cluster.
The overall PDP from $n$ clusters is given by
\begin{equation}
P_d (t) = P_{n}(t)\exp(-\Gamma_{n}/\mu)\delta(t-\Gamma_n)~,
\end{equation}
where $\mu$ is the inter-cluster decay constant given by $\mu\propto c_{\rm d}\Gamma + h/c_{\rm h} + \psi$, with $h$ being the height of the UAV, $c_{\rm d}>1$ and $c_{\rm h}>1$ are constants showing that the inter cluster decay constant is a function of cluster delay and height of UAV, and $\psi\sim\mathcal{N}(0,\sigma_c^2)$ is a zero mean Gaussian random variable. The number of clusters $C_{\rm n}$ is given by $C_{\rm n} \propto c_{\rm e}/h + \gamma$, where  $c_{\rm e}>1$ is a constant dependent on the type of the environment, $h$ is the height of the UAV above ground and $\gamma\sim\mathcal{N}(0,\sigma_N^2)$ is a zero mean Gaussian random variable. The $c_{\rm e}$ will have smaller value for nearby clutter conditions. 

The PDP in case of overlapping clusters can be modeled separately. Overlapping occurs when $\tau_{n-1,m}>\Gamma_{n}-\Gamma_{n-1}$ for some value of $m$. This results in overlapping of the clusters $n$ and $n-1$. In order to find the PDP within each cluster for overlapping case, assume a contribution of duration $\chi$ from the neighboring clusters, where $\chi = \Gamma_{n} - \Gamma_{n-1}$. Let $P_{\rm n_c}(t)$ represents the PDP of overlapping clusters $n$ and $n-1$.
The overall PDP for over-lapping clusters can be modeled as \cite{Saleh}
\begin{align}
P_{\rm n_c}(t) =
\begin{cases}
E(a_{n,m}^2)\exp(-\tau_{n,m}/\beta_n)\delta(t-\tau_{n,m})  \\ \quad \text{if }(\tau_{n-1,m}<\Gamma_{n}-\Gamma_{n-1})\\
E(a_{n,m}^2)\exp(-\tau_{n,m}/\beta_n)\exp(-\chi_j/X)\delta(t-\tau_{n,m}) \\ \quad \text{if }(\tau_{n-1,m}>\Gamma_{n}-\Gamma_{n-1})
\end{cases}\nonumber
\end{align}
while the overall PDP for all the clusters $P_{\rm d_c}(t)$ in overlapping case is given by
\begin{equation}
 P_{\rm d_c}(t)=P_{\rm n_c}(t)\exp(-\Gamma_n/\mu)\delta(t-\Gamma_n)~,
\end{equation}
where $X = E[\beta_n,\beta_{n-1}]$, $\chi_{j}$ represents the $j^{\rm th}$ cluster overlap, $j \in (n,n-1)$. Generally the decay constant for the cluster is much larger than the MPC decay constant, and therefore, the additional exponential term in $P_{\rm n_c}(t)$ will decay much faster and the cluster energies are considered to be uncorrelated~\cite{Saleh}.

Using the PDP information, the mean excess delay and the RMS-DS can be obtained as
\begin{align}
t_{\rm mean} &= \frac{\sum_{\forall t}^{} t T_{\rm s} P_{\rm d}(t)}{\sum_{\forall t}^{}P_{\rm d}(t)}~,\\
t_{\rm rms} &= \sqrt{t_{\rm sq}-t_{\rm mean}^2}~,~
t_{\rm sq} = \frac{\sum_{\forall t}^{} (t T_{\rm s})^2 P_{\rm d}(t)}{\sum_{\forall t}^{}P_{\rm d}(t)}~,
\end{align}
where $t_{\rm mean}$ represents the mean excess delay, $t_{\rm rms}$ represents the RMS-DS obtained from respective PDP. In this case we consider non-overlapping clusters, and the mean number of clusters is represented by $\overline C$.

\subsection{Small Scale Fading}
The small scale amplitudes collected for each MPC at respective delay bins for multiple CIRs follow the Nakagami distribution given by
\begin{equation}
F(y;m,\Omega) = \frac{2m^my^{2m-1}}{\Gamma(m) \Omega^m}\exp\Big(\frac{-my^2}{\Omega}\Big)~,
\end{equation}
where $m$ is the Nakagami shape factor, $\Omega$ is the spread controlling factor, and $\Gamma(m)$ is the Gamma function. Let $Y$ be the random variable given by $Y\sim~Nakagami\big(m,\Omega\big)$; then, the parameters $m$ and $\Omega$ are given by~\cite{kolar}
$m = E^2[Y^2]/{\rm Var}[Y^2]$, $\Omega = E[Y^2]$, where the Nakagami~$m$ factor follows log-normal distribution. The mean $\eta$ and standard deviation $\xi$ of the $m$~factor are given by~\cite{molisch00}
\begin{equation}
\eta = m_{\rm 0} - E[m]\gamma,~ \xi =  v_{\rm 0} - {\rm Var}[m]\gamma~,
\end{equation}
where $m_{\rm 0}$, $v_{\rm 0}$ are the mean and variance of the first component of the clusters, respectively, that are not affected by the delay.


\section{Experimental and Modeling Results}

In this section, based on the measurement results and propagation models described in the earlier two sections, we present our results for UWB AG propagation for various UAV communication scenarios in Fig.~\ref{Fig:Exp_Layout}. First, results for large scale channel measurement and modeling results will be discussed, followed by multipath and small scale results.

\subsection{Large Scale Channel Characterization}

\begin{figure}[!t]
	\centering
	\includegraphics[width=\columnwidth]{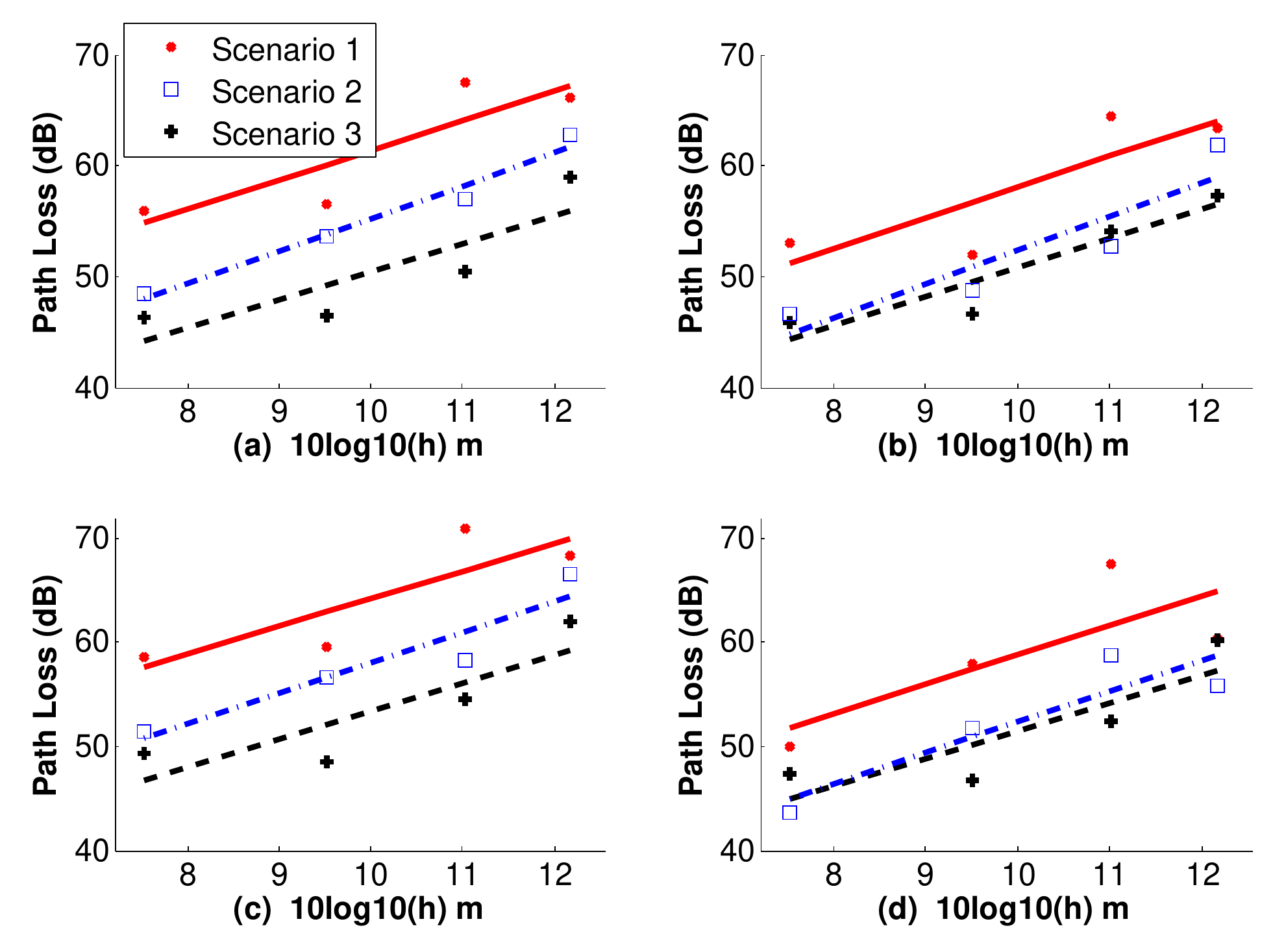}
	\vspace{-0.7cm}
	\caption{Measured path loss versus distance, and linear line fitting to measured data for the three scenarios each in (a) Open area, $v$= 0 m/s, (b) Sub-urban area, $v$ = 0 m/s, (c) Open area, $v$ = 20 m/s, (d) Sub-urban area, $v$ = 20 m/s.}\label{Fig:line_fitting}
\end{figure}

Based on the measured UWB signals at different UAV heights and for the three different scenarios in Fig. 2, the measured path loss versus distance, as well as the linear path loss model fit obtained from (1) for each scenario, are shown in Fig. 3. Results are reported both for open area and sub-urban area, and corresponding parameters for the path loss models are shown in Table~\ref{Table_I} and Table~\ref{Table_II}. It can be observed that the relative motion of the transmitter on UAV with respect to receiver introduces change in antenna's elevation plane pattern, 
which introduces additional path loss and more variance in shadowing. The effect of frequency variance due to Doppler~\eqref{Eq:2} has very little effect in our case due to low velocity of the UAV.

The path loss in Fig.~\ref{Fig:line_fitting} is the highest for scenario~1 with UAV in motion, while it is smallest for scenario~2 (open, LOS case) \eqref{Eq:1}. Path loss is higher for scenario~3 as compared to scenario~2, due to (apart from small change in link distance) capturing of more ground reflections when the receiver is above ground at an optimum height $h_{\rm opt}$. In other words, $h_{\rm opt}$ is the minimum height of the receiver above ground from where we start to capture substantially different behavior of MPCs when compared to the receiver on ground. Another possible reason for the higher path loss in scenario\ 3 is due to more ground absorption of energy and incident angle that is a function of height. In the sub-urban area the path loss is larger than in the open area.

\begin{figure}[!h]
	\centering
	\vspace{-0.2cm} 
	\includegraphics[width=0.9\columnwidth]{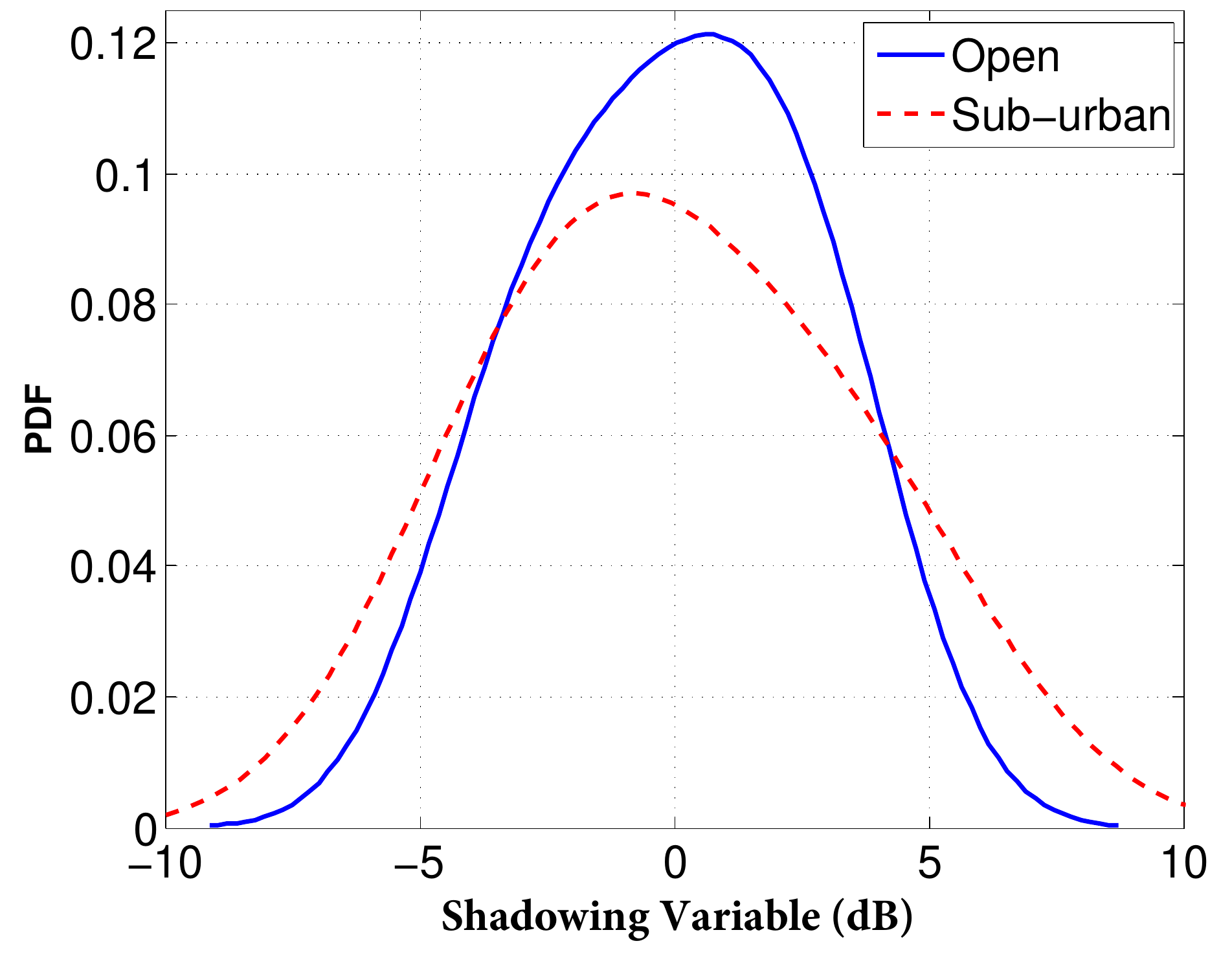}
	\vspace{-0.2cm}
	\caption{PDF of shadowing variable in open area and sub-urban area.}\label{Fig:PDF_Shadowing}
\end{figure}

In Fig.~\ref{Fig:PDF_Shadowing}, the probability density function (PDF) of the variance of path loss is shown for both open and sub-urban areas. Results show that the variance of the path loss is larger for the sub-urban area when compared to the open area measurements.

\begin{align}
{\rm CB} = \frac{1}{5t_{\rm rms}} 
\end{align}

\begin{table}[t]
	\begin{center}
	\caption{Path Loss parameters for ($d = 5.6 \ {\rm m} \ {\rm to} \ 16.5 \ {\rm m}$) open area.}
		\label{Table_I}
		\begin{tabular}{|p{3.5cm}|p{1.1cm}|p{1.15cm}|p{1.1cm}|}
			\hline
			{\ \ \ \ \ \ \ \ \ \ \ \ \ \ \ \textbf{Scenario}} & \ \ \ \ $\alpha$ & $PL_{0}$~(dB)& $\sigma$ (dB)\\
			\hline
			Open, scenario 1, $v$ = 0 mph  & \ $2.6471$ & \ $34.905$ &\ $3.37$ \\
			\hline
			Open, scenario 2, $v$ = 0 mph & \ $2.5418$ & \ $24.9965$ &\ $3.06$ \\
			\hline
			Open, scenario 3, $v$ = 0 mph & \ $2.9442$ & \ $25.8091$ &\ $2.799$ \\
			\hline			
			Open, scenario 1, $v$ = 20 mph  & \ $2.6533$ & \ $34.906$ &\ $4.02$ \\
			\hline
			Open, scenario 2, $v$ = 20 mph  & \ $2.6621$ & \ $24.996$ &\ $3.91$ \\
			\hline
			Open, scenario 3, $v$ = 20 mph  & \ $2.9423$ & \ $25.809$ &\ $3.44$ \\
			\hline			
			
		\end{tabular}
	\end{center}
\end{table}

\begin{table}[!t]
	\centering
\caption{Path Loss parameters for ($d = 5.6 \ {\rm m} \ {\rm to} \ 16.5 \ {\rm m}$) sub-urban area.}\label{Table_II}
	\begin{tabular}{|p{4.0cm}|p{1.05cm}|p{1.2cm}|p{1.1cm}|}
		\hline
		{\ \ \ \ \ \ \ \ \ \ \ \ \textbf{Scenario}} & \ \ \ \ $\alpha$ & $PL_{0}$ (dB)& $\sigma$ (dB)\\
		\hline
		Sub-urban, scenario 1, $v$ = 0 mph & \ $ 2.7601$ & \ $30.4459$ &\ $4.8739$ \\
		\hline
		Sub-urban, scenario 2, $v$ = 0 mph & \ $2.606$ & \ $24.747$ &\ $4.31$ \\
		\hline
		Sub-urban, scenario 3, $v$ = 0 mph & \ $3.0374$ & \ $21.96$ &\ $4.897$ \\
		\hline			
		Sub-urban, scenario 1, $v$ = 20 mph  & \ $2.8350$ & \ $30.446$ &\ $5.3$ \\
		\hline
		Sub-urban, scenario 2, $v$ = 20 mph  & \ $2.667$ & \ $24.833$ &\ $4.96$ \\
		\hline
		Sub-urban, scenario 3, $v$ = 20 mph  & \ $2.961$ & \ $22.73$ &\ $4.71$ \\
		\hline
\end{tabular}

\end{table}


\subsection{Multipath Channel Characterization}

\begin{figure}[!t]
	\centering
	\includegraphics[width=\columnwidth]{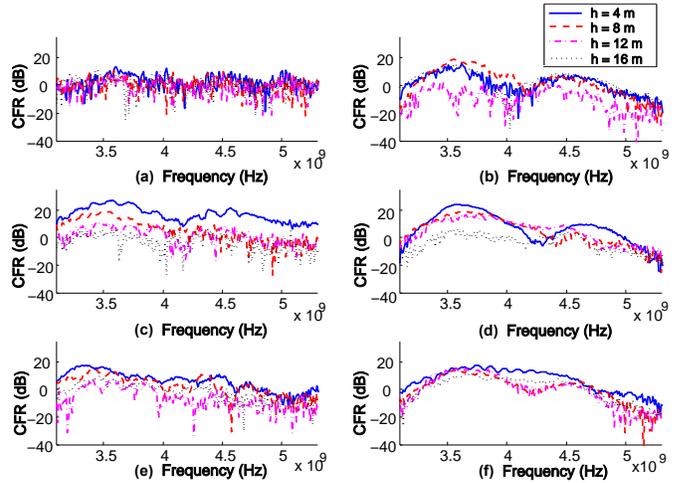}
	\vspace{-0.7cm}
	\caption{CFRs for open and sub-urban areas, for different UAV heights. (a) Open (scenario~1), (b) Sub-urban (scenario~1), (c) Open (scenario~2), (d) Sub-urban (scenario~2), (e) Open (scenario~3), and (f) Sub-urban (scenario~3).}\label{Fig:CFR}
\end{figure}

In order to get insights about the multipath channel characteristics, we have studied various aspects of UAV multipath channel measurements. In Fig.~\ref{Fig:CFR},  channel frequency responses (CFRs) for different scenarios are obtained by taking the FFT of the respective CIRs for different UAV heights. For open area scenario, except for near sharp nulls for any band on the order of $50$~MHz, CFR is fairly flat; however, as the UAV height increases, frequency selective fading is more visible especially in scenario~1, we observe deep fades that become more frequency selective as the UAV height increases.  

\begin{figure}[!t]
	\centering
	\includegraphics[width=\columnwidth]{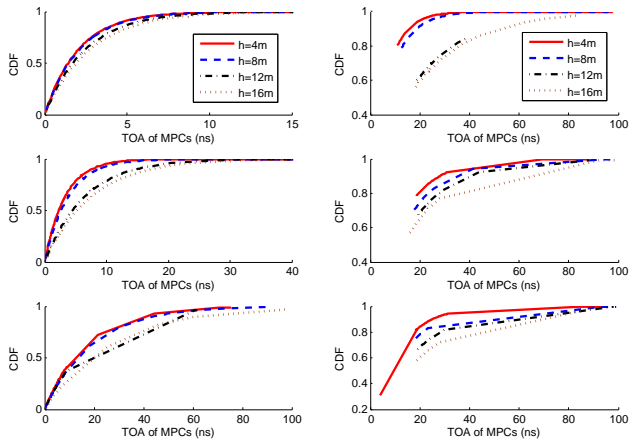}
	\vspace{-0.7cm}
	\caption{CDF versus time of arrival of MPCs. (a) Open (scenario~1), (b) Sub-urban (scenario~1), (c) Open (scenario~2), (d) Sub-urban (scenario~2), (e) Open (scenario~3), and (f) Sub-urban (scenario~3).}\label{Fig:CDF_TOA}
\end{figure}

Fig.~\ref{Fig:CDF_TOA} shows the cumulative distribution function (CDF) for the time of arrival (TOA) of the MPCs for various scenarios. The MPCs are selected by defining an amplitude threshold of $-32.5$~dB, and everything below this threshold is counted as noise and discarded. CDFs in case of open area are closely bound, showing no presence of distant reflectors. In case of sub-urban area the CDFs show more MPCs at larger values of delay, possibly due to arrival of distant MPCs incurring a larger number of reflections. The arrival rate is the highest for scenario~1, due to close by reflections from tree trunk and the branches of the trees.

\begin{figure}[!t]
	\centering
	\includegraphics[width=\columnwidth]{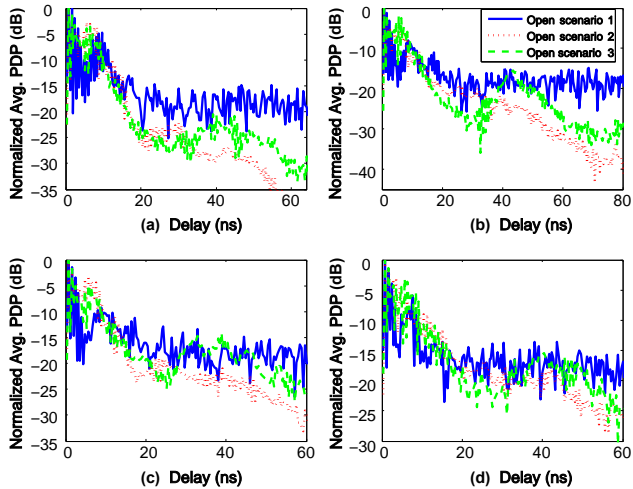}
	\vspace{-0.7cm}
	\caption{PDP for open area for the three scenarios at UAV heights of (a) $4$~m; (b) $8$~m; (c) $12$~m; (d) $16$~m.}\label{Fig:PDP_Open}
\end{figure}

The PDPs for open area are shown in Fig.~\ref{Fig:PDP_Open}
for four different heights of the UAVs and they follow decaying exponential distribution with multiple MPC clusters. There are major power contributions from delayed clusters around $8$~ns and $40$~ns for scenario\ 2 and scenario\ 3. The value of  $\mu$ for scenario 2 and scenario 3 is larger compared to scenario 1 due to larger $\Gamma$. The value of $\Gamma$ that is representation of the clustering delay from MPCs is smallest in case of foliage due to nearby reflections. Similarly due to smaller value of $c_{\rm e}$ for scenario 1 compared to the other two, we have smaller observed $C_{\rm n}$. Similar measurements in sub-urban area (not reported due to space constraints) show that PDP has  more uniform contributions from delayed MPCs forming delayed clusters due to reflections from nearby infrastructure, and larger $\mu$ compared to the open area PDPs. The number of clusters in open and sub-urban areas is approximately the same.

\begin{figure}[!t]
 	\centering
 	\includegraphics[width=\columnwidth]{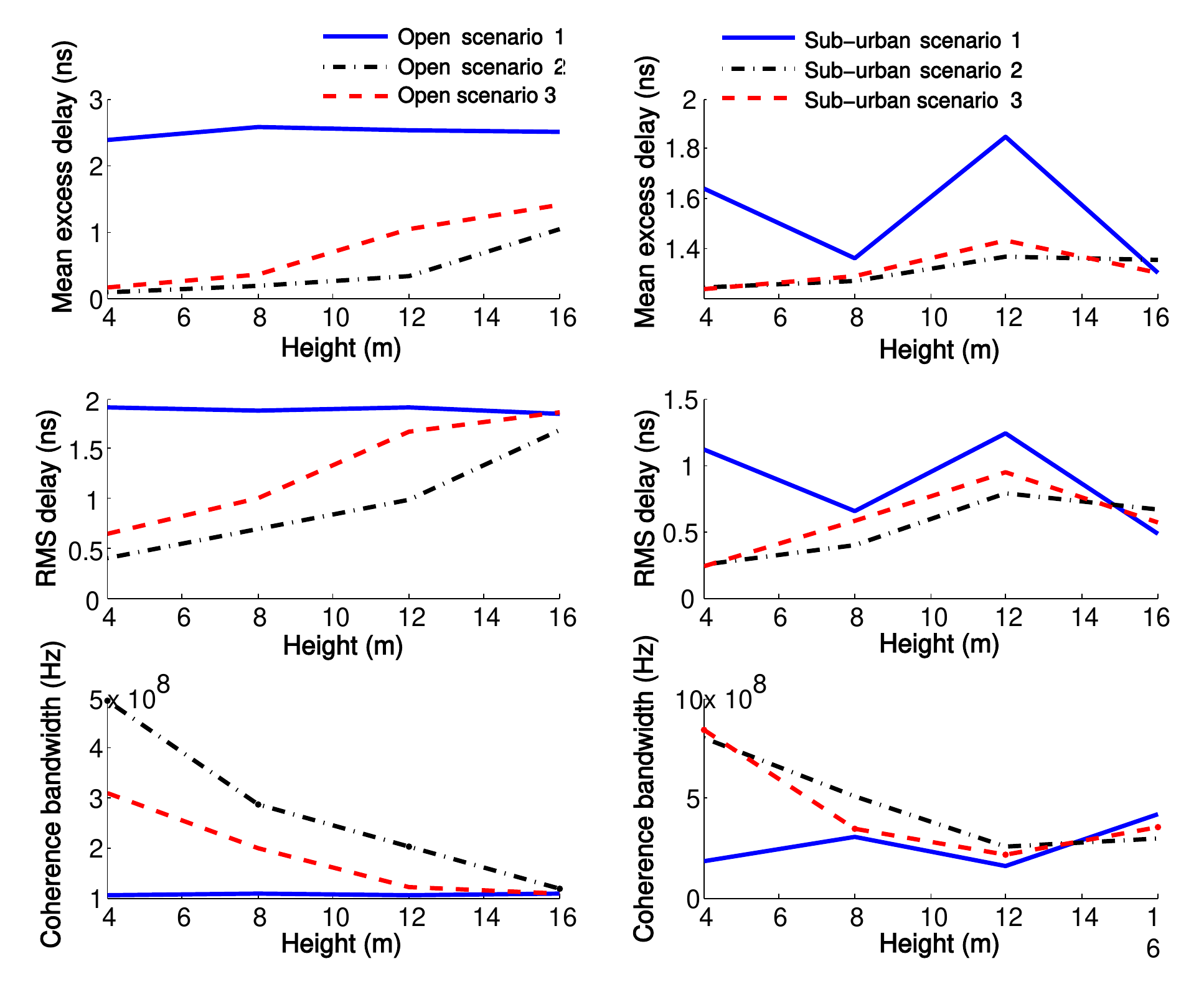}
	\vspace{-0.7cm}
 	\caption{Mean excess delay, RMS-DS and coherence bandwidth for open and sub-urban area at different UAV heights. (a) Open, mean excess delay, (b) Sub-urban, mean excess delay, (c) Open, RMS-DS, (d) Sub-urban, RMS-DS, (e) Open, coherence bandwidth, (f) Sub-urban, coherence bandwidth.}\label{Fig:MED}
 \end{figure}

The mean excess delay, RMS-DS, and coherence bandwidth for open and sub-urban areas are shown in Fig.~\ref{Fig:MED}. For both cases, the mean excess delay and RMS-DS are the highest for scenario~1, and lowest for scenario~2. The coherence bandwidth is calculated using the approximation $1/5t_{\rm rms}$, and it is minimum for foliage conditions. The coherence bandwidth is found to be at least 100 MHz. 

\begin{figure}[!t]
	\centering
	\includegraphics[width=\columnwidth]{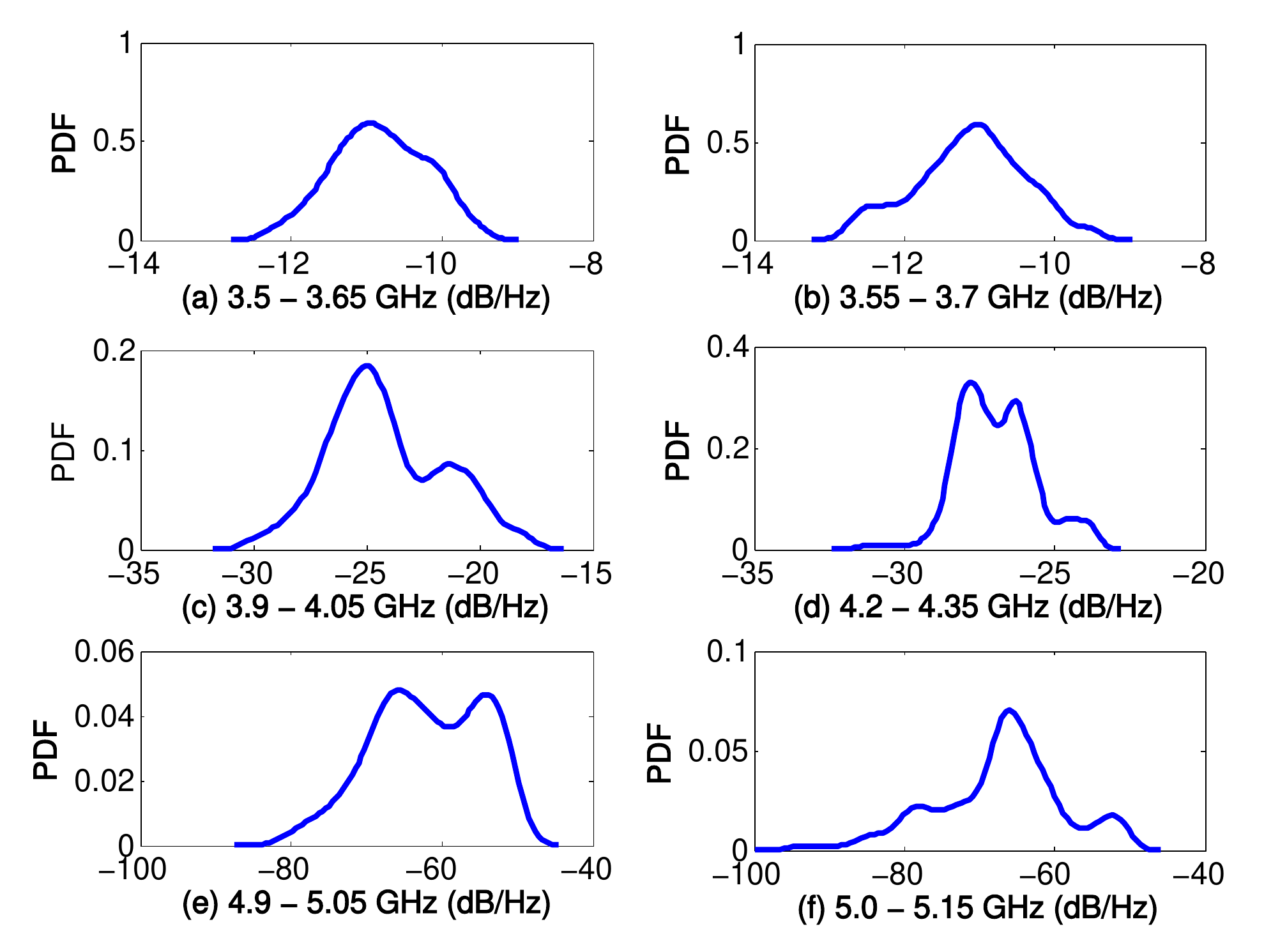}
	\vspace{-0.7cm}
	\caption{PDF of the received signal strength at different frequency sub-bands at a height of 12 m (sub-urban area, scenario~3).}\label{Fig:PDF_Subband}
\end{figure}

In order to study the fading distributions in different frequency sub-bands, we captured the histograms of the received signal strength at different sub-bands (each $150$~MHz wide), over large number of realizations.
Results at a UAV height of $12$~m is shown in Fig.~\ref{Fig:PDF_Subband} for open area scenario. The mean of the PDFs shows a decrease at higher frequencies (due to larger attenuation) and the variances of the PDF shows an increasing trend. 
We also observed that the variances of the PDFs increase with UAV height (results not included).

\subsection{Small Scale Channel Characterization}

Based on the multipath propagation model presented in (5)-(13), we extracted model parameters from open area and sub-urban area measurements, which are reported in Table III. Results show that the mean number of clusters $\overline{C}$ is larger for the sub-urban area, while the inter-cluster decay constant $\mu$ is generally larger for the open area, except for scenario~1.

Moreover, based on the small scale fading model provided earlier in (14), (15), parameters extracted from our UWB AG channel measurements are presented in Table IV for open area and sub-urban area. The mean $\eta$ is smallest and variance $\xi$ is largest for Scenario~1 for both open area and sub-urban area measurements.  In general, variance is larger in sub-urban area compared to open area, due to larger number of scatterers.

\begin{table}[!t]
	\begin{center}
		\caption{UWB UAV channel model parameters for PDP.}\label{Table_IV}
		\begin{tabular}{|p{1.4cm}|p{1.4cm}|p{1.4cm}|p{1.4cm}|}
			\hline
			\textbf{Parameters} & \textbf{Scenario 1} & \textbf{Scenario 2}& \textbf{Scenario 3}\\
			
			\hline
			\multicolumn{4}{|c|}{\textbf{Open area}} \\
			\hline
			$\overline C $  & \ $2.33$ & \ $2.33$ &\ $1$ \\
			\hline
			$\Lambda$ (1/ns) & \ $0.15$ & \ $0.09$ &\ $0.0498$ \\
			\hline
			$\lambda$ (1/ns) & \ $4.34$ & \ $2.210$ &\ $0.532$ \\
			\hline
			$\mu$ (ns) & \ $2.5$ & \ $2.91$ &\ $4.42$ \\
			\hline
			$\beta$ (ns) & \ $0.5$ & \ $0.9069$ &\ $1.21$ \\
			\hline
			\multicolumn{4}{|c|}{\textbf{Sub-urban area}} \\
			\hline	
			$\overline C$  & \ $2.66$ & \ $2.66$ &\ $2.66$ \\
			\hline
			$\Lambda$ (1/ns) & \ $0.789$ & \ $0.0498$ &\ $0.06$ \\
			\hline
			$\lambda$ (1/ns) & \ $0.827$ & \ $0.717$ &\ $0.615$ \\
			\hline
			$\mu$ (ns) & \ $2.63$ & \ $2.77$ &\ $3.03$ \\
			\hline
			$ \beta$ (ns) & \ $0.9$ & \ $1.4$ &\ $1.6$ \\
			\hline

		\end{tabular}
		
	\end{center}
\end{table}

\begin{table}[!t]
	\begin{center}
		
		\caption{UWB UAV channel model parameters for Small scale fading.}\label{Table_III}
		\begin{tabular}{|p{1.4cm}|p{1.4cm}|p{1.4cm}|p{1.4cm}|}
			\hline
			\textbf{Parameters} & \textbf{Scenario 1} & \textbf{Scenario 2}& \textbf{Scenario 3}\\
			\hline
			\multicolumn{4}{|c|}{\textbf{Open area}} \\
			\hline
			$\eta \ (dB)$ & \ $1.36$ & \ $1.67$ &\  $1.45$ \\
			\hline
			$\xi$ & \ $2.19$ & \ $0.64$ &\ $0.79$ \\
			\hline	
			\multicolumn{4}{|c|}{\textbf{Sub-urban area}} \\
			\hline
			$\eta$\ (dB) & \ $1.12 $ & \ $1.58 $ &\ $1.34 $ \\
			\hline
			$\xi$ & \ $2.705$ & \ $1.55$ &\ $1.471$\\
			\hline		
		\end{tabular}
	\end{center}
\end{table}

\section{Concluding Remarks and Applications}

In this work, we conducted extensive UWB AG channel measurements for UAVs for various open and sub-urban scenarios. Based on the empirical data, we have developed statistical channel models for path loss, multipath, and small scale characterization of AG channels, which are shown to match closely with the empirical data. 
Our future work includes extending our propagation model to larger UAV communication distances and heights, in the mmWave and investigate propagation models for air-to-air communication.

Applications of the proposed channel model can be in cognitive radar and environmental sensing systems \cite{5638235}, 3G and 4G cellular networks and millimeter-wave (mmWave) based 5G future cellular networks. In order to reduce outage probability in densely populated areas (e.g. shopping malls and streets), UAVs can act as mobile base stations (BSs) \cite{nadi} to support the additional traffic~\cite{Arvind_VTM_2016}. In case of mmWave communications, the proposed channel model needs to account for angle of arrivals and angle of departures of the MPCs due to extensive use of multiple antennas for mmWave frequencies. Another application is vehicular to infrastructure networks, where the proposed channel model can be used after some modifications to account for the vehicle and UAV velocity.  

\bibliographystyle{IEEEtran}


\end{document}